\documentclass[aps,twocolumn,letterpaper]{revtex4}
\usepackage{amssymb}
\usepackage[dvips]{graphicx}
\usepackage{dcolumn}
\usepackage{bm}
\usepackage{epsfig}

\begin{document}

\title{Evidence for local structural symmetry-breaking in Ca$_{1-x}$Eu$_{x}$B$_{6}$}

\author{H. Martinho$^{1}$}

\author{C. Rettori$^{2}$}

\author{Z. Fisk$^{3}$}

\author{S. B. Oseroff$^{4}$}

\affiliation{$^{1}$Instituto de Pesquisa e Desenvolvimento, UNIVAP, 12244-050,
S\~{a}o Jos\'{e} dos Campos, S\~{a}o Paulo, Brazil}

\affiliation{$^{2}$Instituto de F\'{\i}sica ''Gleb Wataghin'', UNICAMP, 13083-970,
Campinas, S\~{a}o Paulo, Brazil}

\affiliation{$^{3}$Department of Physics, University of California, Davis, CA95616}

\affiliation{$^{4}$San Diego State University, San Diego, CA92182}

\begin{abstract}
In this work we present a systematic Raman Scattering study in the Ca$_{1-x}$%
Eu$_{x}$B$_{6}$ series ($0.00\leqslant x\leqslant 1.00$). Our results are
the first experimental evidence for the occurrence of a local symmetry break
of the crystalline structure in this system. The local symmetry of some
Boron octahedra is tetragonal instead of cubic. This result may explain the
appearance of ferromagnetism in Eu-hexaborades since magnetic ordering is
forbidden in the $O_{h}$ space group of this system.
\end{abstract}

\maketitle

The hexaborides RB$_{6}$ (R = alkaline-earth metal) are one of the most
intensively studied groups of intermetallic compounds presenting a large
variety of physical ground states.\cite{rev} In particular, the family R$%
_{1-x}$A$_{x}$B$_{6}$ (A = magnetic rare-earth) has attracted considerable
interest due to the intriguing connections between their magnetic,\cite{mag}
transport,\cite{transp} and optical properties.\cite{kerr,pereira} These
compounds are cubic with the unit cell belonging to the $O_{h}$ space group.
\cite{cubic} The divalent alkaline metal occupies the central position on a
cube, surrounded by eight B$_{6}$ octahedra at each vertex.\cite{cubic}

In this study we shall focus on the Ca$_{1-x}$Eu$_{x}$B$_{6}$ series with $%
0.00\leqslant x\leqslant 1.00$. EuB$_{6}$ is a ferromagnetic (FM) metal,
characterized by a quite small effective carrier density at high-$T$%
\ ($\sim 10^{-3}$\ per unit cell)\cite{Sullow,carrier1,carrier2} and
presents FM transitions\cite{Sullow} at $T_{C1}=15.3$ K and $T_{C2}=12.5$ K.
The two transitions exhibit pronounced anisotropy and are sensitive to the
application of small magnetic fields.\cite{Sullow} Ferromagnetism is found
to arise from the coupling between the half-filled $4f$ shell of Eu$^{2+}$
ions, whose localized electrons account for the measured magnetic moment of $%
7$$\mu _{B}$ per formula unit.\cite{Sullow,carrier1,carrier2,mB} The FM
transition decreases with Ca content\cite{carrier2,magCa} and are completely
suppressed for $x\lesssim 0.30$.\cite{percolation} Electron microscopy data
indicates the coexistence of insulating Ca- and metallic Eu-rich regions
that percolate at $x\approx 0.30$.\cite{percolation} Moreover, Urbano
\textit{et al} \cite{urbano 1} have recently analyzed the behavior of the Eu$%
^{2+}$ Electron Spin Resonance (ESR) linewidth and reached to the conclusion
that the percolation of the impurity bound states starts already at $%
x\approx 0.15$, i.e., involving next to nearest neighbors.

\begin{figure}[tbh!]
\includegraphics[width=8cm]{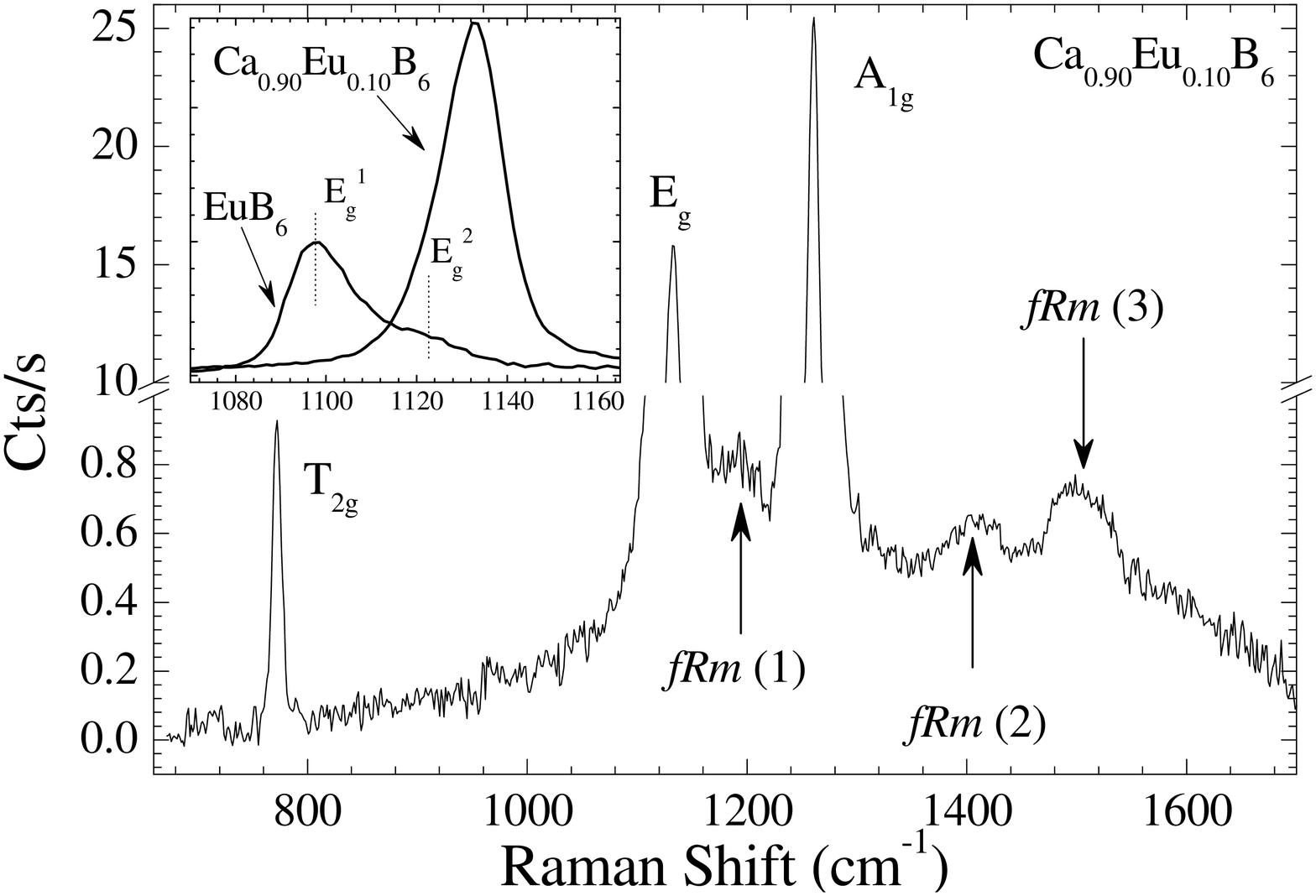}
\caption{Room-$T$ unpolarized RS for Ca$_{0.90}$Eu$_{0.10}$B$_{6}$, showing the expected
phonons of the $O_{h}$ cubic structure as well the forbidden
ones. The inset shows the doublet structure of the $E_{g}$ phonon for EuB$%
_{6}$ and the lack of the doubling for Ca$_{0.90}$Eu$_{0.10}$B$_{6}$. This
lack also occurs for Ca$_{0.95}$Eu$_{0.05}$B$_{6}$ (not shown).}
\label{espectro_representativo}
\end{figure}

The intriguing fact is that group theory prevents the occurrence of FM
ordering in the $O_{h}$ space group of RB$_{6}$.\cite{groupthe} The symmetry
of the lattice must be lowered to one of its tetragonal or orthorhombic
subgroups in order to allow for any FM ordering.\cite{groupthe}
High-resolution $T$-dependent x-ray diffraction measurements by S\"{u}%
llow \textit{et al}\cite{Sullow} failed to observe this symmetry breaking,
indicating that its observation must be extremely subtle. However, recent
ESR $g$-value measurements in the Ca$_{1-x}$Eu$_{x}$B$_{6}$\ series have
revealed that the symmetry of the crystal may be lower than cubic.\cite
{urbano}

The purpose of our systematic Raman Scattering (RS) study in the Ca$_{1-x}$Eu%
$_{x}$B$_{6}$ series ($0.00\leqslant x\leqslant 1.00$) is to find an
evidence for symmetry-breaking by the crystalline structure that could
corroborate the ESR \cite{urbano} results and justify the appearance of FM
ordering in this system. To the best of our knowledge there is no systematic
experimental work concerning\ the $T$-dependence and polarization selection
rules of the RS in hexaborades focusing on the apparent forbidden FM.

For the cubic $O_{h}$ RB$_{6}$ unit cell, the zone-center $T_{2g}$, $E_{g}$,
and $A_{1g}$ phonons are Raman-active. These modes correspond to internal
displacement of B atoms in the B$_{6}$ octahedron.\cite{Udagawa} Previous RS
works\cite{Udagawa,Morke,Yahia,Lemmens,Nyhus1,Nyhus2,Ogita,teredesai} have
reported these modes at $\sim 780$ ($T_{2g}$), $\sim 1150$ ($E_{g}$), and $%
\sim 1260$ ($A_{1g}$) cm$^{-1}$. Besides these phonons, additional modes at $%
\sim 200$ and $\sim 1400$ cm$^{-1}$ were also reported. The former was
conclusively assigned to two-phonon RS\cite{Nyhus2} and the latter has no
clear assignment yet. We believe that a systematic RS study of the $T$%
-dependence and polarization selection rules for these modes may help
clarify the enigma concerning the underlying symmetry problem of the RB$_{6}$
lattice.

\begin{table}[tbh!]
\caption{Polarization selection rules in the $O_{h}$ space group.}
\label{polrules}
\begin{tabular}{|c|c|}
\hline
configuration & Irreducible Representation \\ \hline
$a,b$ & $T_{2g}$ \\
$a,a$ & $A_{1g}+E_{g}+T_{2g}$ \\
$a\prime,a\prime$ & $A_{1g}+E_{g}$ \\
$a\prime,b\prime$ & $A_{1g}+E_{g}$ \\
$c,a$ & $T_{2g}$ \\
$c,c$ & $A_{1g}+E_{g}$ \\
$c\prime,c\prime$ & $A_{1g}+E_{g}+T_{2g}$ \\
$c\prime,a\prime$ & $E_{g}+T_{2g}$ \\ \hline
\end{tabular}
\end{table}

\begin{figure}[th]
\includegraphics[width=6.5cm]{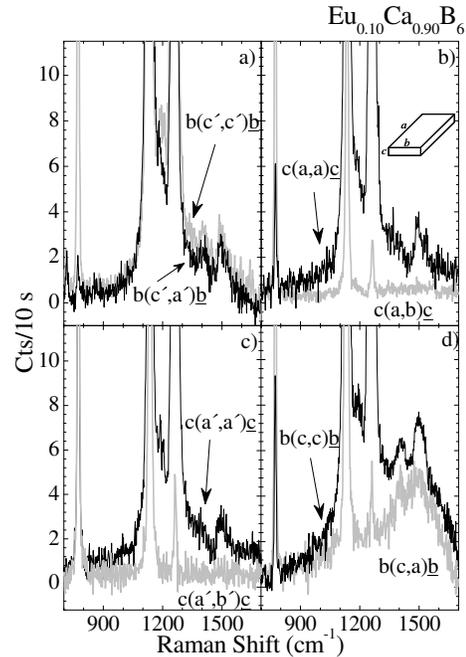}
\caption{Room$-T$ polarized RS spectra for $x=0.10$ in the configurations listed in table
\ref{polrules}. The inset of b) shows the typical crystal shape and the $a,b$ and $c$
axes.} \label{eu10pol}
\end{figure}

Single-crystalline samples of Ca$_{1-x}$Eu$_{x}$B$_{6}$ were grown in a flux of
$99.999\%$ pure Al, using the starting elements Eu, Ca, and B with purities of $99.95\%$,
$99.98\%$ and $99.99\%$, respectively.\cite{cubic} Magnetic and transport
characterization data could be found in ref. \cite {urbano}. The crystal's shapes were
plate-like with typical dimensions of $0.5\times 4\times 6$ mm$^{3}$ (see inset of. Fig.
\ref{eu10pol}). The Raman measurements were carried out using a triple spectrometer
equipped with a LN$_{2}$ CCD detector. The spectra was corrected by the spectrometer
response obtained by measuring the emission of a tungsten lamp and comparing it with the
emissivity of a black-body at same temperature. The $514.5$ nm line of an Ar$^{+}$ ion
laser was used as an excitation source. The laser
power at the sample was kept below $12$ mW on a spot diameter of about $50$ $%
\mu $m. The $x=0.10$ sample was cooled in a cold finger of a He closed-cycle
Displex cryostat. All measurements were made in a near-backscattering
configuration.

\begin{figure}[bh]
\includegraphics[width=8cm]{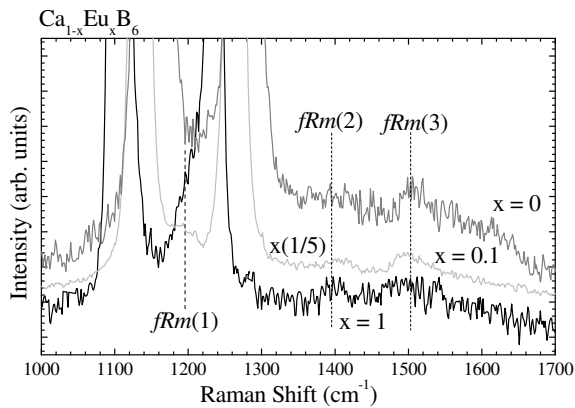}
\caption{Room$-T$ unpolarized RS for $x=0$, $0.10$ and $1$, showing the \textit{fRm} (1)
at $1200$ cm$^{-1}$, \textit{fRm}(2) at $1400$ cm$^{-1}$ and \textit{fRm}(3) at $1500$
cm$^{-1}$.} \label{eu10}
\end{figure}

Fig.\ref{espectro_representativo} presents the RS spectrum at $300$ K for
the $x=0.10$ sample. This is a representative spectrum of the series and
shows the three phonons corresponding to the cubic phase $T_{2g}$, $E_{g}$,
and $A_{1g}$ at $775,1137$ and $1264$ cm$^{-1}$, respectively. Besides these
phonons, extra broad modes at $\sim 1170$, $\sim 1400$ and $\sim 1500$ cm$%
^{-1}$ are clearly observed. To confirm the phononic origin of these extra
modes we used another laser line ($488.0$ nm) to excite the sample and we
found that all the modes were observed at the same position with respect to
the laser line. The only relevant change was that the spectral intensity
response increased by a factor of $5$ when the excitation was the $514.5$ nm
line, indicating a possible resonant behavior. We shall label these extra
modes as forbidden Raman modes \textit{fRm}(1), \textit{fRm}(2) and \textit{%
fRm}(3), respectively. For $x=0$ the \textit{fRm}(1) was not observed.
Besides, we noticed the presence of a shoulder close to the $E_{g}$ phonon
for $x=1$ and $x\geqslant 0.30$. (see inset of Fig.\ref
{espectro_representativo}) This shoulder is always located at $\sim 25$ cm$%
^{-1}$ above the $E_{g}$ phonon frequency. For CaB$_{6}$ and EuB$_{6}$ it
was recently attributed to charge unbalancing in the Boron octahedron.\cite
{Udagawa,Morke,Ogita} It is worth mentioning that for $x=0.05$ and $0.10$
this shoulder is absent, and the $E_{g}$ peak is symmetric, (see inset of
Fig.\ref{espectro_representativo}).

Fig.\ref{eu10} displays the room$-T$ unpolarized RS for the $x=0$, $%
0.10$, and $1$ samples. The spectra show the \textit{fRm}(1), (2) and (3) at $1200$
cm$^{-1}$, $1400$ and $1500$ cm$^{-1}$, respectively. The \textit{fRm}s are present in
all samples, except the \textit{fRm}(1) that
was not observed for $x=0$. We notice that the peaks are more evident for $%
x=0.10$.

In order to get a better insight into the nature of the \textit{fRm}s, it is
helpful to consider their polarization selection rules. The polarized RS
experiments were performed on the $ab$ and $ac$ planes of a monocrystalline
sample with $x=0.10$ at room-$T$. The $a,b$\ and $c$-axes corresponds to the
$[001]$, $[010]$\ or $[100]$\ crystallographic directions.\cite{cubic} The
smaller crystal thickness determines our $c$-axis (see inset of Fig.\ref
{eu10pol}b)). The data are shown in the Fig.\ref{eu10pol}. The expected
selection rules are summarized in Table \ref{polrules}. We notice that the
phonons $A_{1g}$, $E_{g}$, $T_{2g}$ and \textit{fRm}(1) follow the expected
selection rules for the $O_{h}$ space group with \textit{fRm}(1) having the $%
A_{1g}$ symmetry. However, the \textit{fRm}(2) and \textit{fRm}(3) peaks,
both having the same symmetry, do not follow the expected selection rules
for any cubic configuration presented in Table \ref{polrules}. The lack of
cubic symmetry is clearly seen when comparing, e.g., the RS signal in $a,b$
and $c,a$ configurations. They must be equal in the $O_{h}$ space group.

Ogita \textit{et al}\cite{Ogita} observed a broad peak centered at $\sim 1400
$ cm$^{-1}$, probably due to the convolution of \textit{fRm}(2) and \textit{%
fRm}(3). These authors assigned it to an overtone of the $T_{2g}$ phonon. We
disagree with this interpretation for the following reasons: \textit{i}) the
energy of this broad mode is smaller than twice of the $T_{2g}$ phonon
frequency; \textit{ii}) the selection rule for this mode is incompatible
with any combination of the $O_{h}$ irreducible representations (IR),
contrary to what is expected by the direct product $T_{2g}\bigotimes
T_{2g}=A_{1g}+E_{g}+T_{1g}+T_{2g}$;\cite{Koster} and \textit{iii}) their $T$%
-dependence (see below) is incompatible with a two-phonon RS process. Also,
these authores observed a low frequency mode at $\sim 200$ cm$^{-1}$, in
agreement with previous work.\cite{Morke,Nyhus1,Nyhus2} Since this mode was
conclusively assigned to two-phonon RS \cite{Nyhus1} we shall not explore
its selection rules.

\begin{figure}[t!]
\includegraphics[width=8cm]{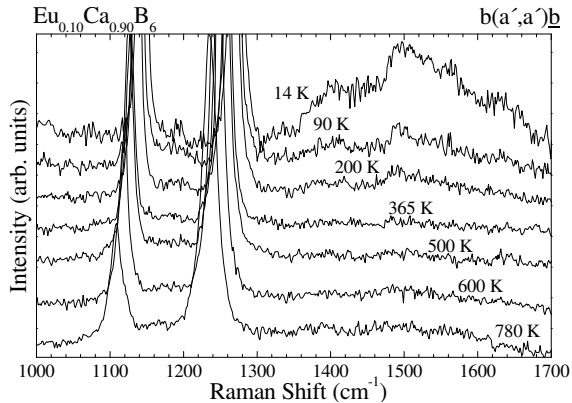}
\caption{ $T-$dependence of the RS spectra for the $x=0.10$ sample at
$a\prime,a\prime$ polarization.} \label{eu10temp}
\end{figure}

Another possible origin for the \textit{fRm}s is disorder-activated
infrared-activeted phonons or combinations between Raman-active and
infrared-active modes. Yahia \textit{et al}\cite{Yahia} measured the RS and
infrared reflectivity for several hexaborides. Comparing the frequencies of
their infrared-active modes with the frequencies of the \textit{fRm}\ we
conclude that there are no infrared mode, neither possible combinations,
that could be assigned to the fRms peaks. Thus, this assumption should be
disregarded.

As a different origin for the \textit{fRm}s we may consider the effect of lowering the
crystal symmetry. In fact, analyzing the\ possible symmetry reduction of the cubic
$O_{h}$ group to a tetragonal or orthorhombic ones, we found that the \textit{fRm}(2) and
\textit{fRm}(3) have the $A$ symmetry of the tetragonal groups. These results may be
understood if one consider that:

\begin{description}
\item[(i)]  the \textit{fRm}(1) as a doublet of the $A_{1g}$ phonon
originated by the charge unbalancing in the B$_{6}$ octahedron, as the $E_{g}
$ doublet, and

\item[(ii)]  the \textit{fRm}(2) and \textit{fRm}(3) as arising from the
local tetragonal symmetry originated by Boron vacancies.
\end{description}

These local symmetry-breaking may allow the FM in this system and justify
the difficulties in resolving large scale symmetry changes with bulk
techniques as x-ray diffraction. Moreover, our $c$-axis corresponds to the $%
[001]$\  tetragonal direction which suggests a cooperative character of the
distortions, presumably associated to a subtle crystal growth habitus. As we
mention above the phonons $A_{1g},E_{g}$, $T_{2g}$ and \textit{fRm}(1) do
not obey the selection rules for the tetragonal symmetry. This suggests that
the overall crystal distortion may be very small.

\begin{figure}[bh]
\includegraphics[width=7cm]{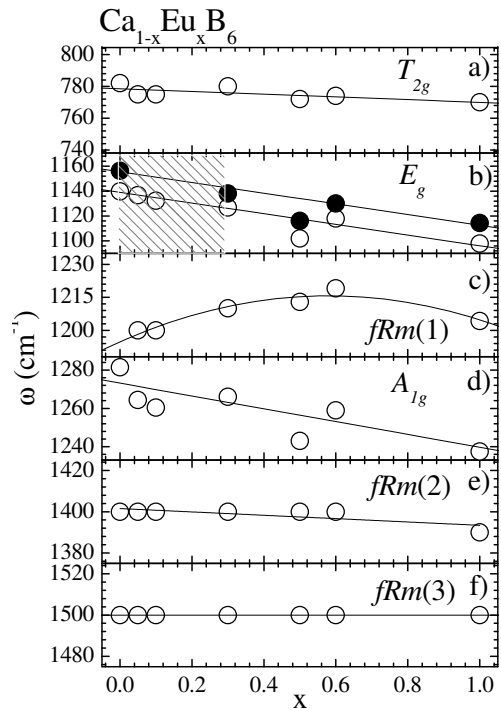}
\caption{Room$-T$ phonon frequencies as function of $x$. Closed circles represents the
high energy $E_{g}$ doublet counterpart and the solid lines are guide for the eyes.}
\label{eu10xfreq}
\end{figure}

Fig.\ref{eu10temp} presents the $T$-dependence of the RS spectra for $x=0.10$
. While the \textit{fRm}(1) keeps its intensity almost constant, the \textit{%
fRm}(2) and \textit{fRm}(3) present a strong low-$T$ intensity increase. The
integrated area of \textit{fRm}(2)+ \textit{fRm}(3) increases by a factor $6$
between $780$ and $14$ K (not shown). For $T>400$ K, the peak's intensities
are nearly constant. We should mention that this $T$-dependence is not
clearly understood yet. We notice that the $T$-dependence of the phonon
frequencies and linewidths for $x=0.10$ displayed the expected thermal
lattice anarmonicity behavior.\cite{Menendez}

Fig.\ref{eu10xfreq} shows the phonon frequencies at $300$ K as a function of
Eu content $x$. For the $O_{h}$ phonons and \textit{fRm}(1) , the
frequencies were obtained from fitting the spectra to Lorentzian lineshapes%
while for the \textit{fRm}(2) and (3) it was obtained at the position of the maximum
intensity. The solid lines are guides for the eyes. The linear softening as function of
$x$ for the $O_{h}$ phonons (Fig.\ref {eu10xfreq} a), b) and d)) is consistent with the
corresponding increase of the lattice parameter that varies between $\sim 4.14$ \AA\ for
$x=0$\cite {Ogita} and $\sim 4.18$ \AA\ for $x=1.00$.\cite{Sullow} The lack of $E_{g}$
doubling can be seen in the $0<x<0.30$ interval and it is represented by the dashed area
in Fig.\ref{eu10xfreq} b). Electron microscopy and resistivity results by Wigger
\textit{et al}\cite{percolation} indicate that in this region there is a phase separation
between insulating Ca-and metallic Eu-rich regions that percolate at $x\sim 0.30$. This
may reinforce the interpretation about the origin of the $E_{g}$\ doubling that relies in
the charge unbalance on the B$_{6}$ octahedron. The absence of the $E_{g}$\ doubling in
this interval may be related to the lack of long range coherence that screens the
doubling of the $E_{g}$\ mode. Increasing the doping beyond $x\sim 0.30$\ will
reintroduce the long range coherence in the charge distribution enabling again the
doubling of the $E_{g}$\ mode. Similar doubling is seen in CaB$_{6}$ due to the charge
unbalance caused by defects in the sample, which may introduce carriers by
doping.\cite{urbano3} Another interesting result is the nearly parabolic behavior of the
\textit{fRm}(1) with a maximum around $x\sim 0.45$. Following our previous idea about the
origin of this mode, this behavior may also depend on the competition between
insulating-Ca and metallic-Eu-rich regions.

In conclusion, our systematic RS results show the appearance of three \textit{fRm}s
in the Ca$_{1-x}$Eu$_{x}$B$_{6}$ series that does not belong to the expected set of
Raman-active phonons of the $O_{h}$ cubic group.
After analyzing different possibilities we suggested that the \textit{fRm}%
(1) is a doublet of the $A_{1g}$ phonon, caused by the charge unbalance in
the B$_{6}$ octahedron, similar to the $E_{g}$ doublet. Besides, the \textit{%
fRm}(2) and \textit{fRm}(3) peaks arise from the local cubic
symmetry-breaking originating from Boron vacancies that induces a local
tetragonal environment in the B$_{6}$ octahedron. A cooperative behavior of
these local tetragonal distortions may explain the appearance of FM in
Eu-hexaborates and a tetragonal axis along the smaller crystal dimension.
Moreover, their local character would probably justify the difficulties that
bulk techniques, such as x-ray diffraction. had in observing this small
scale symmetry changes. We should mention that evidences for local
distortions of the Boron octahedra were observed in SmB$_{6}$ hexaborates by
Raman\cite{Nyhus2} and ESR.\cite{sturm} It appears that local distortions
play an important role in describing the physical properties of hexaborades.

\textbf{Acknowledgments}. This work was supported by the Brazilian Agencies
CNPq and FAPESP. We would like to thank Dr. S. Lance Cooper for fruitful
discussions and critical reading of the manuscript.

\end{document}